\documentclass[12pt,twoside]{article}
\usepackage{fleqn,espcrc1}

\usepackage{epsfig}
\usepackage{rotate}

\title{Chiral Corrections to Baryon Masses Calculated within Lattice
QCD}

\author{Anthony W. Thomas\footnote{athomas@physics.adelaide.edu.au}, 
	Derek B. Leinweber\footnote{dleinweb@physics.adelaide.edu.au}, 
        Kazuo Tsushima\footnote{ktsushim@physics.adelaide.edu.au},  and 
        Stewart V. Wright\footnote{swright@physics.adelaide.edu.au}}

\begin{document}

\maketitle

\vspace{-4.cm}
\hfill ADP-99-38/T375
\vspace{4.cm}

\noindent
Department of Physics and Mathematical Physics\\
and Special Research Centre for the Subatomic Structure of Matter,\\
University of Adelaide, Australia 5005\\

\begin{abstract}
Consideration of the analytic properties of pion-induced
baryon self energies leads to new functional forms for the extrapolation
of light baryon masses.  These functional forms reproduce the leading
non-analytic behavior of chiral perturbation theory, the correct
non-analytic behavior at the $N \pi$ threshold and the appropriate
heavy-quark limit. They involve  
only three unknown parameters, which may be obtained by fitting
lattice QCD data. Recent dynamical fermion results from CP-PACS
and UKQCD are extrapolated using these new functional forms. We also
use these functions to probe the limit of applicability of chiral
perturbation theory.
\end{abstract}

\section{Introduction}

Chiral symmetry requires that the nucleon mass has the form
\begin{displaymath}
m_{N}(m_{\pi}) = m_{N}(0) + \alpha m_{\pi}^2 + \beta m_{\pi}^3 +
\gamma m_{\pi}^{4} \ln m_{\pi} + \ldots \; ,
\end{displaymath}
for small $m_{\pi}$, where $m_{N}(0)$, $\alpha$, $\beta$, and $\gamma$
are functions of the strong coupling constant $\alpha_{s}(\mu)$.
Recent work \cite{Leinweber:1999ig} has shown that using
physical insights from chiral perturbation theory and heavy quark
effective theory one can derive new functional forms which describe
the extrapolation of light baryon masses as functions of the pion mass
$(m_{\pi})$.  These forms are applicable beyond the chiral
perturbative regime and have been compared successfully with
predictions from the Cloudy Bag Model \cite{Thomas:1984kv} and recent
dynamical fermion lattice QCD calculations.

\section{Analyticity}

\label{Analyticity-sec}

\begin{figure}[t]
\centering{\
\epsfig{file=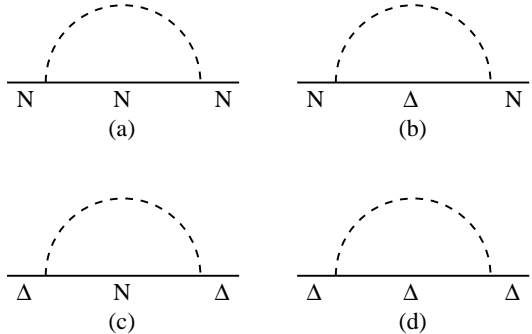,width=7cm}}
\caption{One-loop pion induced self energy of the nucleon and the
delta.
\label{SE-fig}}
\end{figure}

By now it is well established that chiral symmetry is dynamically
broken in QCD and that the pion is almost a Goldstone boson. It is
strongly coupled to baryons and therefore plays a 
significant role in the $N$ and $\Delta$ self energies. In the
limit where the baryons are heavy, the pion-induced self energies of the
$N$ and $\Delta$, to one loop, are given by the processes
shown in Fig.~\ref{SE-fig}(a--d).  We label these by $\sigma_{NN}$,
$\sigma_{N\Delta}$, $\sigma_{\Delta N}$, and
$\sigma_{\Delta\Delta}$. Note that we have restricted the intermediate
baryon states to those most strongly coupled, namely the 
$N$ and $\Delta$ states. Other intermediate states are suppressed by
the baryon form factor describing the extended nature of baryons.

The leading non-analytic contribution (LNAC) of these self energy
diagrams is associated with the infrared behavior of the
corresponding integrals -- i.e., the behavior as the loop momentum
$k\rightarrow 0$. As a consequence, it should not depend on
the details of a high momentum cut-off, or form factor. In
particular, it is sufficient for
studying the LNAC to evaluate the self energy integrals using a simple
sharp cut-off, $u(k)=\theta (\Lambda -k)$ as the choice of form
factor. The explicit forms of the self energy
contributions for $\sigma_{NN}$, 
$\sigma_{N\Delta}$ and so on are given in \cite{Leinweber:1999ig}.
Moreover, there is little phenomenological difference
between this step function and the more natural dipole, provided one
can tune the cut-off parameter $\Lambda$. The self energies involving
transitions of $N \to \Delta$ or $\Delta \to N$ are characterized by a
branch point at $m_\pi = \Delta M$.

\subsection{Chiral Limit\label{chirlim}}

The leading non-analytic (LNA) terms are those which correspond to the
lowest order non-analytic functions of $m_{q}$ -- i.e., odd powers or
logarithms of $m_{\pi }$. By expanding the expressions given in
\cite{Leinweber:1999ig}, we find that the LNA contributions to the
nucleon/delta masses are
in agreement with the well known results of $\chi$PT
\cite{HBchiPT,Lebed94}.

Of course, our concern with respect to lattice QCD is not so much the
behavior as $m_{\pi }\rightarrow 0$, but the extrapolation from
high pion masses to the physical pion mass. In this context the branch
point at $m_{\pi }^{2}=\Delta M^{2}$ is at least as important as
the LNA behaviour near $m_{\pi }=0$.

\subsection{Heavy Quark Limit}

Heavy quark effective theory suggests that as $m_{\pi }\rightarrow
\infty$ the quarks become static and hadron masses become
proportional to the quark mass. In this spirit, corrections are
expected to be of order $1/m_{q}$ where $m_{q}$ is the heavy quark
mass. Thus we would expect the pion induced self energy to vanish as
$1/m_{q}$ as the pion mass increases. The presence of a fixed
cut-off $\Lambda$ acts to suppress the pion induced self energy
for increasing pion masses. While some $m_{\pi }^{2}$ dependence in
$\Lambda$ is expected, this is a second-order effect and does not
alter this qualitative feature. Indeed, in the large
$m_{\pi }$ limit of the equations, we find that they
tend to zero at least as fast as $1/m_{\pi}^{2}$.

The agreement with both the chiral limit and expected behaviour in the
heavy quark limit suggests the following
functional form for the extrapolation of the nucleon mass \cite{Leinweber:1999ig}:
\begin{equation}
\label{N-form-eqn}
M_{N}=\alpha _{N}+\beta _{N}m_{\pi }^{2}+\sigma _{NN}(m_{\pi },\Lambda
)+\sigma _{N\Delta }(m_{\pi },\Lambda )\, .
\end{equation}

\section{Lattice Data Analysis}

\label{lattice-fit-sec}

We consider two independent lattice simulations of the $N$ and
$\Delta$ masses from CP-PACS \cite{CP-PACSlight} and UKQCD
\cite{Allton:1998gi}.  Both of these use improved actions to study baryon
masses in full QCD with two light flavours.
We find that the two data sets
are consistent, provided one allows the parameters introducing the
physical scale to float within systematic errors of 10\%.

\begin{figure}[t]
\centering{\
\rotate{\epsfig{file=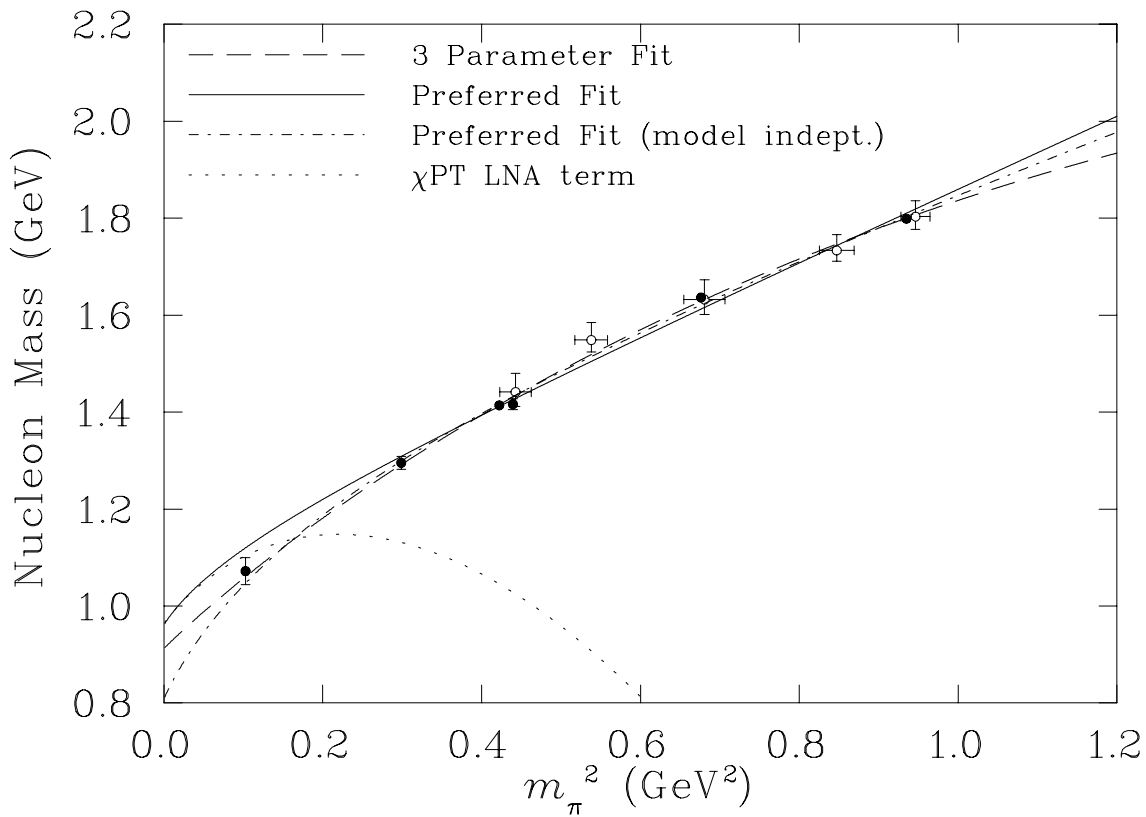,height=15cm}}
\caption{A comparison between phenomenological fitting functions for
the mass of the nucleon.  The dotted curve corresponds to using
Eq.\ (\ref{Naive-eqn}) with $\gamma$ set equal to
the value known from $\chi$PT.  The three
parameter fit (dashed) corresponds to letting $\gamma$
vary as an unconstrained fit parameter. The solid and dash-dot curves
correspond to  our preferred fit of the functional form of Eq.\
(\ref{N-form-eqn}) with $\Lambda$ from the CBM and as a fit parameter
respectively. The lattice data from are
CP-PACS (solid) and UKQCD (open), each with a 5\% scale change.
\label{3fits-N-fig}}}
\end{figure}

We begin by considering the functional form suggested in Section
\ref{Analyticity-sec} with the cut-off $\Lambda$ fixed to the
value determined by fitting CBM calculations.  This is shown as
the solid curve in Fig. \ref{3fits-N-fig}.  In order to perform
model independent fits (i.e. with $\Lambda$ unconstrained), it is
essential to have lattice simulations at light quark masses
approaching $m^{2}_{\pi }\sim 0.1$ GeV$^{2}$.  This fit is illustrated
by the dash-dot curve.

Common practice in the lattice community to use a polynomial
expansion for the mass dependence of hadron masses.  Motivated by
$\chi$PT the lowest odd power of $m_\pi$ allowed is $m_\pi^3$:
\begin{equation}
\label{Naive-eqn}
M_{N}=\alpha +\beta m_{\pi }^{2}+\gamma m_{\pi }^{3}
\end{equation}
The result of such a fit for the $N$ is shown in the dashed curve of
Fig. \ref{3fits-N-fig}. The coefficient of the $m_{\pi }^{3}$ term,
which is the leading non-analytic term in the quark mass, in the three
parameter fit is $-0.761$.  This disagrees with the coefficient of
$-5.60$ known from $\chi$PT (which is correctly incorporated in
Eq.\ (\ref{N-form-eqn}), the solid and dash-dot curves) by almost an order of
magnitude. This clearly indicates the failings of such a simple
fitting procedure.

\section{Summary}
\label{discussion-sec}

In the quest to connect lattice measurements with the physical regime,
we have explored the quark mass dependence of the $N$ and $\Delta$
baryon masses using arguments based on analyticity and heavy quark
limits. We have determined a method to access quark masses beyond the
regime of chiral perturbation theory. This method reproduces the
leading non-analytic behavior of $\chi$PT and accounts for the
internal structure for the baryon under investigation.  We find that
the leading non-analytic term of the chiral expansion dominates from
the chiral limit up to the branch point at $m_{\pi}=\Delta M \simeq
300$ MeV, beyond which $\chi$PT breaks down. The predictions of the
CBM, and two-flavour dynamical-fermion lattice QCD results, are
succinctly described by the formulae derived in
\cite{Leinweber:1999ig}. The curvature around $m_{\pi }=\Delta M$,
neglected in previous extrapolations of the lattice data, leads to
shifts in the extrapolated masses of the same order as the departure
of lattice estimates from experimental measurements.

\subsection*{Acknowledgments}

This work was supported in part by the Australian Research Council.

\end{document}